\documentstyle[twocolumn,aps]{revtex}

\begin{document}
\draft
\title{Fluctuation-dissipation considerations and damping models for
ferromagnetic materials}
\author{Vladimir L. Safonov and H. Neal Bertram}
\address{Center for Magnetic Recording Research, \\
University of California - San Diego, \\
9500 Gilman Drive, La Jolla, CA 92093-0401}
\maketitle

\begin{abstract}
The role of fluctuation-dissipation relations (theorems) for the
magnetization dynamics with Landau-Lifshitz-Gilbert and Bloch-Bloembergen
damping terms are discussed. We demonstrate that the use of the
Callen-Welton fluctuation-dissipation theorem that was proven for
Hamiltonian systems can give an inconsistent result for magnetic systems
with dissipation.
\end{abstract}


\section{Introduction}

The study of linear stochastic magnetization dynamics is of great importance
in applications to nano-magnetic devices and ultra-thin films. In general,
gyromagnetic magnetization motion around an effective field is randomly
forced by fluctuations on spins by means of interaction with a thermal bath
(phonons, magnons, conduction electrons, impurities, etc.).
Fluctuation-dissipation relations give a very useful tool to find a
correspondence between the dynamic variable fluctuations, temperature and
dissipation for a given magnetic dynamic system (see, e.g., \cite{brown}, 
\cite{white},\cite{safbertbook},\cite{smitharnett},\cite{BertJinSaf}).

The most frequently used fluctuation-dissipation relations are known as the
Callen-Welton fluctuation-dissipation theorem (FDT) \cite{callen-welton}, 
\cite{LandauLifshitz}. There are two integrals that express this theorem.
The first one gives the correlation function of the dynamic variables in
terms of an integral of the system susceptibility. The second form relates
the correlation function of the applied noise to an integral of the inverse
susceptibility. This theorem is proved for rather general assumptions (we
will discuss them later) and therefore appears to be very attractive for
general problems.\ The aim of this paper is to discuss the inapplicability
of the second relation to magnetization dynamics in ferromagnetic systems.
In particular, we demonstrate that Callen-Welton FDT gives mathematical
inconsistency and can not provide an argument to discriminate between
different damping models.

\section{Fluctuation-dissipation theorem}

The Callen-Welton fluctuation-dissipation theorem is proved under very
general assumptions for {\it Hamiltonian systems} of an arbitrary type (see,
e.g., \cite{tatarskii},\cite{klimontovich}). These systems have no internal
dissipation at all. The energy ${\cal V}$\ of the external perturbation
acting on the system is taken in the form: 
\begin{equation}
{\cal V}=-%
\mathop{\displaystyle\sum}%
\limits_{j}x_{j}(t)f_{j}(t),  \label{interaction}
\end{equation}
where $x_{j}(t)$ is the $j$-th component of the dynamic variable and $%
f_{j}(t)$ is the $j$-th component of the external alternating field. All
dynamic variables are assumed to be equal to zero in the absence of the
perturbation and the linear responses are defined by

\begin{equation}
x_{j}(t)=%
\mathop{\displaystyle\sum}%
\limits_{k}%
\displaystyle\int %
\limits_{-\infty }^{t}K_{jk}(t-\tau )f_{k}(\tau )d\tau ,  \label{response}
\end{equation}
where $K_{jk}(t)$ is a memory function that depends on the properties of the
dynamic system \cite{comment1}. A \textquotedblleft thermal
bath\textquotedblright\ in this theorem is modeled as a set of periodic
external fields, which are absorbed by the dynamic system. On the other
hand, the external fields stimulate radiation from the system and cause a
loss of energy. For a periodic external field

\begin{equation}
f_{j}(t)=\frac{1}{2}\left[ f_{0j}e^{-i\omega t}+f_{0j}^{\ast }e^{i\omega t}%
\right]  \label{random field}
\end{equation}
one can rewrite Eq. (\ref{response}) in the form:

\begin{equation}
x_{j}(t)=\frac{1}{2}%
\mathop{\displaystyle\sum}%
\limits_{k}\left[ \chi _{jk}(\omega )f_{0k}e^{-i\omega t}+\chi _{jk}^{\ast
}(\omega )f_{0k}^{\ast }e^{i\omega t}\right]  \label{dynamic variable}
\end{equation}
where

\begin{equation}
\chi _{jk}(\omega )=%
\displaystyle\int %
\limits_{0}^{\infty }K_{jk}(t)\exp (i\omega t)dt  \label{susceptibility}
\end{equation}
is the susceptibility tensor.

The change of the total energy ${\cal E}$ (which includes the perturbation
energy ${\cal V}$) in the dynamic system that plays an important part in
this theory is expressed as

\begin{equation}
\frac{d{\cal E}}{dt}=-%
\mathop{\displaystyle\sum}%
\limits_{j}x_{j}(t)\frac{df_{j}(t)}{dt}.  \label{change of energy}
\end{equation}
Averaging this equation over the period of the external fields (\ref{random
field}) and taking into account Eq. (\ref{dynamic variable}), one obtains
the following expression for the dissipated power:

\begin{equation}
Q=\frac{i\omega }{4}%
\mathop{\displaystyle\sum}%
\limits_{j,k}\left[ \chi _{jk}^{\ast }(\omega )-\chi _{kj}(\omega )\right]
f_{0j}f_{0k}^{\ast }.  \label{dissipation}
\end{equation}
\qquad

In the derivation both the absorbed and radiated averaged powers are usually
expressed quantum mechanically in terms of transitions probabilities (per
unit time) between infinitely narrow energy levels, i.e., no linewidth or
damping is taken into account. Temperature is introduced by the
thermodynamic weights of the energy levels. This gives a thermal averaging
over a thermal bath. Summing up over frequencies $\omega $, one can obtain
the balance between absorption and irradiation. This balance expressed in
terms of dynamic susceptibility, gives the Callen-Welton FDT. In the
classical limit it can be written as:

\begin{equation}
\langle x_{j}(\tau )x_{k}(0)\rangle =\frac{k_{B}T}{2\pi }\int_{-\infty
}^{\infty }\frac{\chi _{kj}(\omega )-\chi _{jk}^{\ast }(\omega )}{i\omega }%
e^{-i\omega \tau }d\omega .  \label{f1}
\end{equation}
Here $k_{B}$ is the Boltzman constant, and $T$ is the temperature. $\langle
\ldots \rangle $ denotes both thermal averaging and averaging over random
phases in the periodic fields.

A second form of the Callen-Welton FDT is also introduced:

\begin{equation}
\langle f_{j}(\tau )f_{k}(0)\rangle =\frac{k_{B}T}{2\pi }\int_{-\infty
}^{\infty }\frac{\chi _{kj}^{\ast -1}(\omega )-\chi _{jk}^{-1}(\omega )}{%
i\omega }e^{-i\omega \tau }d\omega .  \label{f2}
\end{equation}
This form results from the reversibility of the linear relation 
\begin{equation}
x_{j}(\omega )=\chi _{jk}(\omega )f_{k}(\omega )  \label{x to f}
\end{equation}
to 
\begin{equation}
f_{j}(\omega )=\chi _{jk}^{-1}(\omega )x_{k}(\omega ).  \label{f to x}
\end{equation}

It should be noted that the first (\ref{f1}) and second (\ref{f2}) relations
can be interpreted differently. Namely, in the first case we have a
reasonable correspondence between the correlations of dynamic variables and
the susceptibility of the dynamic system. Changing the dynamic system
properties, and hence the susceptibility, we change the correlations of the
dynamic variable. A specific external noise mechanism is excluded and the
role of the thermal bath is included implicitly by assuming a system in
thermal equilibrium. This gives a linear dependence on the temperature $T$.

On the other hand, in the second case the noise correlations are associated
with the inverse susceptibility of the dynamic system. In other words, two
principally different physical characteristics, the noise, which results
from a thermal bath dynamics and the susceptibility, which describes just
the properties of the dynamic system, should correspond to each other. From
Eq. (\ref{f2}) it follows that by changing properties of the dynamic system,
one can change the noise correlations.

However, physically it seems impossible in general that the dynamic
susceptibility determines the noise variance. It is not clear that the
correspondence between (\ref{x to f}) and (\ref{f to x}) applies for general
random processes. This follows from the fact that the integral (\ref%
{response}) depends on the form of stochastic integration because $%
f_{k}(\tau )$ are random variables. For example, the Ito and Stratonovich
approaches to stochastic integration have no general relationship between
each other in the case of multiplicative noise \cite{gardiner}.

In applications to systems with dissipation, the Callen-Welton
fluctuation-dissipation theorem may be used just as an approximation, which
must be validated. For some dissipative systems the Callen-Welton FDT gives
reasonable estimates. It covers, for example, the Einstein relation for
Brownian motion and the classical Nyquist formula for voltage fluctuations.
However, these particular cases do not prove that the theorem is applicable
for any linear system with dissipation. Some criticism of the Callen-Welton
FDT is given by Klimontovich \cite{klimontovich},\cite{klimontovich1}.

\section{FDT and magnetic damping}

A commonly used theoretical approach to magnetic dynamics, which is purely
phenomenological, is based on the Landau-Lifshitz equation \cite{landau}, or
its modification in the Gilbert form \cite{gilbert}. Local and isotropic
phenomenological damping terms and corresponding local random fields are
assumed to describe a \textquotedblleft thermal bath\textquotedblright\ in a
magnetic material. On the other hand, a microscopic approach predicts
non-local field fluctuations \cite{bertramwang} and magnetization dynamics
in the form of the Bloch-Bloembergen equations: the Landau-Lifshitz-Gilbert
equations occur only for uniaxial symmetry \cite{SafBertNoise},\cite%
{mechanisms}.

Recently Smith \cite{smith0} using the second form of the Callen-Welton
theorem (\ref{f2}) claimed that fluctuation-dissipation relations provide a
means for discriminating between alternative phenomenological magnetic
damping models.\ Here we demonstrate that, by using Eq. (\ref{f2}), the
fluctuation-dissipation relations for stochastic dynamics with
Landau-Lifshitz-Gilbert damping and Bloch-Bloembergen damping are
inconsistent.

For simplicity we shall consider a single-domain particle with the magnetic
energy ${\cal E}$ in the vicinity of equilibrium:

\begin{equation}
\frac{{\cal E}}{M_{s}V}=\frac{H_{x}}{2}m_{x}^{2}+\frac{H_{y}}{2}%
m_{y}^{2}-h_{x}m_{x}-h_{y}m_{y}.  \label{energyE}
\end{equation}
Here $m_{x}\ $and $m_{y}$ are the transverse components of the unit vector $%
{\bf m}\ $oriented along the magnetization,\ $M_{s}$ is the saturation
magnetization, $V$ is the volume of the sample, $H_{x}$ and $H_{y}$ are the
Kittel stiffness fields (in general $H_{x}\neq H_{y}$, for example, in a
thin film). For a general stochastic problem $h_{x}(t)$ and $h_{y}(t)$ are
random variables arising from the interaction with a thermal bath. To
illustrate the utilization of Callen-Welton FDT one assumes that $h_{x}(t)$
and $h_{y}(t)$ are equivalent to the external alternating fields $f_{j}$ in
Eq.(1) that leads to the forms of Eqs.(8) and (9).

The linearized Landau-Lifshitz-Gilbert equation can be written in the form:

\begin{eqnarray}
\frac{1}{\gamma }\left( \stackrel{\cdot }{m}_{x}+\alpha \stackrel{\cdot }{m}%
_{y}\right) &=&-H_{y}m_{y}+h_{y}(t),  \nonumber \\
\frac{1}{\gamma }\left( \stackrel{\cdot }{m}_{y}-\alpha \stackrel{\cdot }{m}%
_{x}\right) &=&H_{x}m_{x}-h_{x}(t),  \label{Gilb}
\end{eqnarray}
or, equivalently as

\begin{eqnarray}
\frac{1}{\widetilde{\gamma }}\stackrel{\cdot }{m}_{x} &=&-\alpha
H_{x}m_{x}-H_{y}m_{y}+h_{y}(t)+\alpha h_{x}(t),  \nonumber \\
\frac{1}{\widetilde{\gamma }}\stackrel{\cdot }{m}_{y} &=&-\alpha
H_{y}m_{y}+H_{x}m_{x}-h_{x}(t)+\alpha h_{y}(t),  \label{LLG}
\end{eqnarray}
where $\widetilde{\gamma }=\gamma /(1+\alpha ^{2})$, $\gamma $ is the
gyromagnetic ratio and $\alpha $ is a dimensionless damping parameter. From
Eq. (\ref{Gilb}) one can obtain the inverse susceptibility

\begin{equation}
\chi _{jk}^{-1}(\omega )=\frac{1}{\gamma M_{s}V}\left( 
\begin{array}{cc}
-i\alpha \omega +\gamma H_{x} & i\omega \\ 
-i\omega & -i\alpha \omega +\gamma H_{y}%
\end{array}
\right) ,  \label{Gsuscep}
\end{equation}
where $h_{j}(\omega )=\chi _{jk}^{-1}(\omega )m_{k}(\omega )M_{s}V$; $%
m_{k}(\omega )M_{s}V$ is the $k$-th component of the magnetic moment.
Substituting Eq. (\ref{Gsuscep}) and its Hermitian conjugate into Eq. (\ref%
{f2}), the second form of Callen-Welton FDT can be expressed as

\begin{equation}
\left( 
\begin{array}{cc}
\langle h_{x}(\tau )h_{x}(0)\rangle & \langle h_{x}(\tau )h_{y}(0)\rangle \\ 
\langle h_{y}(\tau )h_{x}(0)\rangle & \langle h_{y}(\tau )h_{y}(0)\rangle%
\end{array}
\right) =\frac{2k_{B}T\alpha }{\gamma M_{s}V}\left( 
\begin{array}{cc}
1 & 0 \\ 
0 & 1%
\end{array}
\right) \delta (\tau ).  \label{GilbNoise}
\end{equation}

The case of linearized Bloch-Bloembergen magnetization dynamics is described
by

\begin{eqnarray}
\frac{1}{\gamma }\stackrel{\cdot }{m}_{x} &=&-\frac{1}{\gamma T_{2}}%
m_{x}-H_{y}m_{y}+h_{y}(t),  \nonumber \\
\frac{1}{\gamma }\stackrel{\cdot }{m}_{y} &=&-\frac{1}{\gamma T_{2}}%
m_{y}+H_{x}m_{x}-h_{x}(t),  \label{BBD}
\end{eqnarray}
where $T_{2}$ is the relaxation time. From this equation we can obtain the
following inverse susceptibility:

\begin{equation}
\chi _{jk}^{-1}(\omega )=\frac{1}{\gamma M_{s}V}\left( 
\begin{array}{cc}
\gamma H_{x} & i\omega -T_{2}^{-1} \\ 
-i\omega +T_{2}^{-1} & \gamma H_{y}%
\end{array}
\right) .  \label{BBsuscep}
\end{equation}
Application of Eq. (\ref{f2}) yields FDT:

\begin{equation}
\left( 
\begin{array}{cc}
\langle h_{x}(\tau )h_{x}(0)\rangle & \langle h_{x}(\tau )h_{y}(0)\rangle \\ 
\langle h_{y}(\tau )h_{x}(0)\rangle & \langle h_{y}(\tau )h_{y}(0)\rangle%
\end{array}
\right) =\frac{k_{B}T}{\gamma M_{s}V}\left( 
\begin{array}{cc}
0 & -1 \\ 
1 & 0%
\end{array}
\right) \frac{{\rm sgn}(\tau )}{T_{2}}.  \label{FDT-LLLtoBB}
\end{equation}
Note that the difference between Eq. (\ref{GilbNoise}) and (\ref{FDT-LLLtoBB}%
) lies in the tensor on the right hand side of each equation. Because of the
diagonal tensor, Eq. (\ref{GilbNoise}) implies no correlations between $%
h_{x} $ and $h_{y}$. On the other hand, Eq. (\ref{FDT-LLLtoBB}) does show
such correlations. One can argue that the Bloch-Bloembergen form of damping
is not physical due to the non-diagonal form of (\ref{FDT-LLLtoBB}).
However, as we show below, there is an inconsistency in the use of the
second form of the Callen-Welton FDT.

Let us now consider the uniaxial case when both dynamics should coincide. We
examine the dynamics of a soft micromagnetic particle (no anisotropy) in an
external magnetic field $H_{0}$ and small damping approximation $\alpha \ll
1 $ (the most interesting case), when terms $\sim \alpha ^{2}$ can be
neglected. Thus $H_{x}=H_{y}=H_{0}$, and the Landau-Lifshitz-Gilbert
equation (\ref{LLG}) is reduced to:

\begin{eqnarray}
\frac{1}{\gamma }\frac{dm_{x}}{dt} &=&-\alpha
H_{0}m_{x}-H_{0}m_{y}+h_{y}(t)+\alpha h_{x}(t),  \nonumber \\
\frac{1}{\gamma }\frac{dm_{y}}{dt} &=&-\alpha
H_{0}m_{y}+H_{0}m_{x}-h_{x}(t)+\alpha h_{y}(t).  \label{Mequa1}
\end{eqnarray}
So far as $h_{x}(t)$ and $h_{y}(t)$ represent two independent random fields,
their linear combinations are also random. We can consider the following
orthogonal transformation:

\begin{eqnarray}
\widetilde{h}_{y}(t) &=&\frac{h_{y}(t)+\alpha h_{x}(t)}{\sqrt{1+\alpha ^{2}}}%
\simeq h_{y}(t)+\alpha h_{x}(t),  \nonumber \\
-\widetilde{h}_{x}(t) &=&\frac{-h_{x}(t)+\alpha h_{y}(t)}{\sqrt{1+\alpha ^{2}%
}}\simeq -h_{x}(t)+\alpha h_{y}(t).  \label{orth}
\end{eqnarray}
Thus, random fields $\widetilde{h}_{x}(t)$ and $\widetilde{h}_{y}(t)$ are
independent and Eq. (\ref{Mequa1}) can be rewritten in the form:

\begin{eqnarray}
\frac{1}{\gamma }\frac{dm_{x}}{dt} &=&\alpha H_{0}m_{x}-H_{0}m_{y}+%
\widetilde{h}_{y}(t),  \nonumber \\
\frac{1}{\gamma }\frac{dm_{y}}{dt} &=&\alpha H_{0}m_{y}+H_{0}m_{x}-%
\widetilde{h}_{x}(t).  \label{BB}
\end{eqnarray}
From Eqs. (\ref{GilbNoise}) and (\ref{orth}) one can easily calculate pair
correlations for $\widetilde{h}_{x}(t)$ and $\widetilde{h}_{y}(t)$ (the
terms $\sim \alpha ^{2}$ are neglected):

\begin{equation}
\left( 
\begin{array}{cc}
\langle \widetilde{h}_{x}(\tau )\widetilde{h}_{x}(0)\rangle & \langle 
\widetilde{h}_{x}(\tau )\widetilde{h}_{y}(0)\rangle \\ 
\langle \widetilde{h}_{y}(\tau )\widetilde{h}_{x}(0)\rangle & \langle 
\widetilde{h}_{y}(\tau )\widetilde{h}_{y}(0)\rangle%
\end{array}
\right) =\frac{2kT\alpha }{\gamma M_{s}V}\left( 
\begin{array}{cc}
1 & 0 \\ 
0 & 1%
\end{array}
\right) \delta (\tau ).  \label{BBthroughG}
\end{equation}

Note that the Eq. (\ref{BB}) is an exact mathematical analog of the
Bloch-Bloembergen equation (\ref{BBD}) ($1/T_{2}=\alpha \gamma H_{0}$) for
the transverse magnetic components with random fields $\widetilde{h}_{x}(t)$
and $\widetilde{h}_{y}(t)$. Thus, we see that the fluctuation-dissipation
relations (\ref{BBthroughG}) and (\ref{FDT-LLLtoBB}) are principally
different in the case when they must coincide (Gilbert and Bloch-Bloembergen
dynamics are the same). The second form of the Callen-Welton FDT can not be
used to verify the validity of one form of damping versus another.

The origin of the inconsistency is in the use of the second form (\ref{f2})
of Callen-Welton FDT, applied to systems with dissipation
(Landau-Lifshitz-Gilbert and Bloch-Bloembergen equations). Without
dissipation ($\alpha =0$) this inconsistency disappears. It should be
pointed out that Eqs. (\ref{Mequa1}) and (\ref{BB}) describe stochastic
magnetization dynamics with $\left\langle h_{x}(t)\right\rangle
=\left\langle h_{y}(t)\right\rangle =\left\langle m_{x}(t)\right\rangle
=\left\langle m_{y}(t)\right\rangle =0$ in accordance with thermodynamics.

Note that for no damping the dynamic susceptibility of the
Landau-Lifshitz-Gilbert equation (\ref{Gsuscep}) and the Bloch-Bloembergen
equation (\ref{BBsuscep}) are identical in form. In the stochastic case with
random fields $h_{x}(t)$ and $\left\langle h_{y}(t)\right\rangle $ at first
glance Eqs. (14) and (17) appear to be different. However, we have shown
that by a simple transformation (21) both equations are equivalent in the
case of uniaxial anisotropy.

Application of the first FDT (\ref{f1}) to both Landau-Lifshitz-Gilbert and
Bloch-Bloembergen stochastic dynamics does not give such a strong
inconsistency as does the second form. Using (\ref{f1}) with (\ref{Gsuscep}%
), we obtain the following averages for the Gilbert dynamics:

\begin{eqnarray}
&&\left( 
\begin{array}{cc}
\langle m_{x}(\tau )m_{x}(0)\rangle & \langle m_{x}(\tau )m_{y}(0)\rangle \\ 
\langle m_{y}(\tau )m_{x}(0)\rangle & \langle m_{y}(\tau )m_{y}(0)\rangle%
\end{array}
\right)  \label{f1G} \\
&=&\frac{\gamma \alpha k_{B}T}{\pi M_{s}V}\int_{-\infty }^{\infty }\frac{%
d\omega }{|D_{G}(\omega )|^{2}}e^{-i\omega \tau }  \nonumber \\
&&\times \left( 
\begin{array}{cc}
\omega ^{2}(1+\alpha ^{2})+(\gamma H_{y})^{2} & -i\omega \gamma (H_{x}+H_{y})
\\ 
i\omega \gamma (H_{x}+H_{y}) & \omega ^{2}(1+\alpha ^{2})+(\gamma H_{x})^{2}%
\end{array}
\right) ,  \nonumber
\end{eqnarray}
where $D_{G}(\omega )=\omega _{0}^{2}-\omega ^{2}(1+\alpha ^{2})-i\alpha
\omega \gamma (H_{x}+H_{y})$ and $\omega _{0}^{2}=\gamma ^{2}H_{x}H_{y}$ is
the ferromagnetic resonance frequency.

On the other hand, the use of (\ref{f1}) with (\ref{BBsuscep}) gives:

\begin{eqnarray}
&&\left( 
\begin{array}{cc}
\langle m_{x}(\tau )m_{x}(0)\rangle & \langle m_{x}(\tau )m_{y}(0)\rangle \\ 
\langle m_{y}(\tau )m_{x}(0)\rangle & \langle m_{y}(\tau )m_{y}(0)\rangle%
\end{array}
\right)  \label{f1BB} \\
&=&\frac{\gamma T_{2}^{-1}k_{B}T}{\pi M_{s}V}\int_{-\infty }^{\infty }\frac{%
d\omega }{i\omega |D_{BB}(\omega )|^{2}}e^{-i\omega \tau }  \nonumber \\
&&\left( 
\begin{array}{cc}
i2\omega \gamma H_{y} & \omega _{0}^{2}+\omega ^{2}+T_{2}^{-2} \\ 
-\omega _{0}^{2}-\omega ^{2}-T_{2}^{-2} & i2\omega \gamma H_{x}%
\end{array}
\right) ,  \nonumber
\end{eqnarray}
where $D_{BB}(\omega )=\omega _{0}^{2}-\omega ^{2}+T_{2}^{-2}-2i\omega
T_{2}^{-1}$.

In general, magnetic correlations (\ref{f1G}) and (\ref{f1BB}) differ from
each other. However, in the most important case of the noise power ($\tau =0$%
) both (\ref{f1G}) and (\ref{f1BB}) are reduced to

\begin{equation}
\left( 
\begin{array}{cc}
\langle m_{x}^{2}(0)\rangle & \langle m_{x}(0)m_{y}(0)\rangle \\ 
\langle m_{y}(0)m_{x}(0)\rangle & \langle m_{y}^{2}(0)\rangle%
\end{array}
\right) =\frac{k_{B}T}{M_{s}V}\left( 
\begin{array}{cc}
H_{x}^{-1} & 0 \\ 
0 & H_{y}^{-1}%
\end{array}
\right) .  \label{thermo}
\end{equation}
This equation is completely consistent with thermodynamics, namely, with
energy equipartion:

\begin{equation}
\frac{\langle {\cal E}\rangle }{2}=\frac{M_{s}VH_{x}}{2}\langle
m_{x}^{2}\rangle =\frac{M_{s}VH_{y}}{2}\langle m_{y}^{2}\rangle =\frac{k_{B}T%
}{2}.  \label{equipartition}
\end{equation}

\section{Discussion}

Other forms of the fluctuation-dissipation relations, which are similar to
the first Callen-Welton FDT, have been derived by Kubo \cite{kubo} for the
permeability of a dynamic system and by White \cite{white} for the
susceptibility of a general magnetic system. A standard method to study
stochastic dynamics in systems with dissipation is the Langevin approach 
\cite{klimontovich},\cite{langevin}. Application of the Langevin approach
utilizing specific dissipation mechanisms can be found in \cite{bertramwang}%
, \cite{SafBertNoise},\cite{mechanisms}.

The analysis of damping in magnetic systems has had a long history (e.g., 
\cite{sparks},\cite{Akhiezer}). Each spin wave (magnon) interacts with a
so-called, thermal bath, which consists of magnons, phonons, conduction
electrons, impurities, etc. Analyzing the processes of relaxation, one can
find the spin-wave damping (see, e.g., \cite{sparks}) and the corresponding
thermal noise. The microscopic stochastic differential equation (SDE) is
shown to be of the Langevin form for a damped harmonic oscillator for a wide
class of relaxation processes \cite{mechanisms}. Note that the collective
description works even in the case of local interactions. A local defect or
impurity perturbs the band structure of the magnetic crystal (see, e.g., 
\cite{white}) and affects the spin-wave spectrum, whose imaginary part is
responsible for the damping of collective magnetic excitations.

The occurrence of delocalized damping has been directly demonstrated in the
problem of two coupled magnetic grains with local thermal baths \cite%
{bertramwang}. Stochastic forces are uncorrelated in the spin-wave
coordinates, but become correlated when the SDE's are expressed in the
original magnetization coordinates. Because of the system interactions, even
though the coupling to the thermal bath may be purely local, there is no
physical requirement that the stochastic forces in the magnetization
coordinates be uncorrelated.

In summary, we have argued that the second form of the Callen-Welton
fluctuation dissipation theorem does not correctly apply to damped systems,
has inconsistencies in its application and thus can not distinguish between
relaxation models.

\section{Acknowledgment}

We thank H. Suhl, R. M. White and X. Wang for helpful discussions. This work
was partly supported by matching funds from the Center for Magnetic
Recording Research at the University of California - San Diego and CMRR
incorporated sponsor accounts.


\begin{references}
\bibitem{comment1} It should be noted that in the case when $f_{k}(\tau )$
are random variables, the responses $x_{j}(t)$ depend on the definition of
the stochastic integral (see, e.g., \cite{gardiner}).
\end{references}
\end{document}